\\
Title: On Surface Plasmon Damping in Metallic Nanoparticles
Authors: Armen Melikyan and Hayk Minassian (State Engineering University of Armenia, Yerevan, Armenia)
Comments: 6 pages;   E-mail address:  physdep@seua.am
 Subj-class: Mesoscopic Systems and Quantum Hall Effect; Optics
\\

Two possible mechanisms of surface plasmon (SP) oscillations damping in metallic nanoparticles (MNPs), not connected with electron-phonon interaction are investigated theoretically: a) the radiation damping of SP, b) resonant coupling of SP oscillations with electronic transitions in matrix. It is shown that the radiation damping rate is proportional to the number of electrons in MNP and therefore  this channel of energy outflow from MNP becomes essential for relatively large particles. The investigation of second mechanism shows that the rate of SP oscillations energy leakage from MNP dos not depend on particle size and  is fully determined by the optical characteristics of the matrix. It is demonstrated that for very small MNPs of 3-5 nm size, where the strong 3D size quantization effect suppresses the electron-phonon interaction, the resonance coupling in certain cases provides an effective energy outflow.
\\

**Introduction**

The MNPs in transparent matrices show spectrally selective optical absorption due to collective oscillations and interband transitions [1]. The surface plasmon (SP) frequency of MNPs, unlike bulk samples and films, falls into the visible range due to geometry of nanoparticle and dielectric properties of matrix [1-7], and this important peculiarity is of great interest for a range of optical applications [1,8-13]. When the SP resonance is well separated from other transitions, the central problem in applications is the behavior of SP, i.e. the damping and dephasing of plasma oscillations.

Although there are many observations of  SP resonance in various MNPs [14,15], the mechanisms of SP oscillations damping and dephasing are not completely studied theoretically and identified experimentally. The basic damping mechanism is connected with the electron-phonon interaction processes. It has been noted many times however [16,17], that such processes in systems with strong 3D size quantization  are suppressed because of  kinematical restrictions imposed on the energy-momentum conservation laws. Indeed, in MNPs of size of 3-5 nm and less the electron level spacing is comparable or larger than the Debye energy, and therefore the electron-phonon interaction may become ineffective. On the other hand in several experiments with such small MNPs the absorption line broadening has been observed [17-19]. This means that besides the electron-phonon scattering  other damping  mechanisms should be discussed as well in order to interpret the observed SP line broadenings.

In this paper we consider the radiation damping mechanism and the SP oscillation energy resonance transfer to the matrix. Both these mechanisms are not conditioned by the electron-phonon interaction, and thus they are not suppressed by the size quantization effects.

**1. Radiation damping of SP oscillations.**

One of the possible mechanism of decrease of oscillation energy in MNP is the emission of electromagnetic waves - so called  radiation damping. To the best of our knowledge the radiation damping of SP oscillations is considered in literature only phenomenologically, that is the damping rate is expressed in terms of susceptibility [20]. We propose fully microscopical calculation, which for spherical MNPs allows to determine the SP damping rate dependence on all parameters including the dielectric constant of matrix.

The instant value of SP oscillation energy $W = (N_e m_e v^2)/2$  can be presented as

$$W = \frac{(N_e m_e)\omega_{sp}^2 d^2}{2e^2}, \qquad (1)$$

where $N_e$ is the number of electrons, v is the instant amplitude of velocity of oscillations with frequency $\omega_{sp}$, $d$ is the dipole moment of one electron, $e$ is its charge and $m_e$ is the electron mass. The Poynting vector of radiation field is $c(\vec{E} \times \vec{H})/4\pi$. Substituting $\frac{c}{n} = \frac{c}{\sqrt{\varepsilon_1}}$ for $c$ and $\varepsilon_1 \vec{E}$ for $\vec{E}$ we obtain the energy losses per unit time owing to emission of radiation in media with dielectric constant $\varepsilon_1$ [21]

$$\frac{dW}{dt} = -\frac{\omega_{sp}^4 (N_e d)^2 \sqrt{\varepsilon_1}}{3c^3}. \qquad (2)$$

Comparing Eqs (1) and (2) with rate equation for $W$

$$\frac{dW}{dt} = -\frac{1}{\tau_{rad}} W, \qquad (3)$$

we obtain

$$\frac{1}{\tau_{rad}} = \frac{2e^2}{3mc^3} \omega_{sp}^2 \sqrt{\varepsilon_1} N_e, \qquad (4)$$

where $\tau_{rad}$ is the radiation damping time of SP oscillations, $\omega_p^2 = 4\pi n_e e^2/m$ determines the plasma oscillations frequency in bulk, $n_e$ is the electron concentration. The frequency $\omega_p$ is connected with SP frequency of spherical nanoparticle embedded in a matrix with real part of dielectric constant $\varepsilon_1$ by the expression $\omega_{sp} = \omega_p/\sqrt{1 + 2\varepsilon_1}$ (see [1]). The expression (4) can be also presented in form, which demonstrates the volume dependence of the damping rate

$$\frac{1}{\tau_{rad}} = \frac{2\omega_p^4}{9c^3} \frac{\sqrt{\varepsilon_1}}{1 + 2\varepsilon_1} R_0^3, \qquad (5)$$

where $R_0$ is the MNP radius. As it follows from (4) and (5) the radiation damping rate is proportional to the number of electrons in MNP or its volume, thus, this channel of SP damping plays an important role for relatively large MNPs.

In fact to have a complete description of SP oscillation damping process we should include into Eq. (3) also the term corresponding the electron-phonon mechanism which will lead to the following form of this equation

$$\frac{dW}{dt} = -\left(\frac{1}{\tau_{rad}} + \frac{1}{\tau_{e-ph}}\right) W. \qquad (6)$$

The most important feature of Eq (4), that is the proportionality of the damping rate to the number of electrons in nanoparticle, arises due to coherent oscillations of many particles. The additional reasonings confirming this dependence can be presented as follows. As the size of nanoparticle is much smaller than the wavelength corresponding to SP radiation, the interaction between electrons via radiation field in the nanoparticle can be described by involving the reaction field [22], i.e. the field created by the particle at distances much smaller than the wavelength

$$\vec{E}_r(t) = \frac{2}{3c^3} \dddot{\vec{d}}(t). \tag{7}$$

All the electrons coherently oscillate with the same phase, so that the resultant field acting on one of $N_e \gg 1$ electrons will be $\frac{2N_e}{3c^3}\dddot{\vec{d}}$ and the equation of motion of this electron takes the form

$$\ddot{\vec{d}} + \omega_{sp}^2 \vec{d} + \frac{2N_e e^2}{3mc^3} \dddot{\vec{d}} = 0. \tag{8}$$

In deriving the expression (7) it was assumed, that the reaction field is much less than the quasielastic force $-\omega_{sp}^2 \vec{d}$. Thus from zero order equation $\ddot{\vec{d}} + \omega_{sp}^2 \vec{d} = 0$ we estimate the reaction field and obtain $\vec{E}_r = -\frac{2\omega_{sp}^2}{3c^3}\dot{\vec{d}}$, and the Eq. (8) takes the form

$$\ddot{\vec{d}} + \frac{2e^2 N_e \omega_{sp}^2}{3mc^3} \dot{\vec{d}} + \omega_{sp}^2 \vec{d} = 0. \tag{9}$$

It follows from (9), that the radiation damping rate is indeed proportional to the total number of oscillating electrons $N_e$ in MNP (see (4)). Thus we have an obvious manifestation of so called superradiance [23], the effect, which arises due to interaction of electrons via radiation field when the size of a system of radiating electrons is much smaller than the wavelength.

It follows from (5), that the radiation damping time of 20 nm diameter Au MNP embedded in glass matrix with $\varepsilon_1 = 4$ is approximately $2 \cdot 10^{-14} s$, which is close to the electron relaxation time [24]. The variation of diameter of MNP around 20 *nm* according to (4) will lead to the change of energy leakage rate. Since at this range of sizes $\tau_{e-ph}$ does not depend on the size, the measurements will allow to distinguish the radiation damping mechanism of SP. It should be mentioned, that in case of larger particles the SP oscillations can not be excited, as the penetration depth of the radiation in visible region is about 15 nm.

It would be interesting to carry out the measurements with larger MNPs at very law temperatures, when the electron-phonon interaction is substantially weaker but the radiation damping rate dominates due to the cubic dependence on MNP size and can exceed the rate of heating of the lattice due to the energy transfer from electrons to phonons.

### 2. Resonant coupling of SP oscillations with matrix.

The other mechanism of the energy outflow from MNP is conditioned by the resonant coupling of SP oscillations with the matrix which reveals itself when the frequency of oscillations is close to that of the electron transitions in matrix. Here we consider the case of so called weak coupling, when the matrix absorption spectrum width is very large [25], then the SP oscillation energy irreversibly decreases owing to the leakage into the matrix.

Consider a MNP located at the origin of the reference frame. The dipole moment $\vec{D}(t) = N_e \vec{d}(t) = \vec{D}_0 \cos\omega_{sp} t$ of the oscillating electrons in the nanoparticle creates at point $\vec{r}$ an alternating electric field $\vec{E}(\vec{r}) = \vec{E}_\omega(\vec{r})\cos\omega_{sp}(t - r/v)$, where $v$ is the speed of light in the matrix, $\omega_{sp}$ is determined by (5), and the field strength amplitude has the form

$$\vec{E}_\omega(\vec{r}) = \frac{\vec{D}_0 - 3\vec{n}(\vec{D}_0\vec{n})}{\varepsilon_1 r^3}, \tag{10}$$

$\vec{n}$ is the unit vector in direction of radius vector $\vec{r}$ [22]. This field in its turn creates a polarization $\vec{P}(\vec{r})$

$$\vec{P}(\vec{r}) = Re[(\chi_1 + i\chi_2)\vec{E}_\omega(\vec{r})\exp i\omega_{sp}(t - r/c)] = \\ [\chi_1 \cos\omega_{sp}(t - r/c) - \chi_1 \sin\omega_{sp}(t - r/c)]\vec{E}_\omega(\vec{r}), \tag{11}$$

where $\chi_1$ and $\chi_2$ are the frequency dependent real and imaginary parts of the matrix susceptibility correspondingly. Thus the dielectric constant is

$$\varepsilon(\omega) = \varepsilon_1(\omega) + i\varepsilon_2(\omega) = 1 + 4\pi[\chi_1(\omega) + i\chi_2(\omega)],$$

the imaginary part being much smaller than the real part $\varepsilon_2 \ll \varepsilon_1$. The energy absorbed in matrix per unit time per unit volume averaged over the oscillation period is

$$\langle \vec{E}(\vec{r})\cdot\dot{\vec{P}}(r)\rangle = -\frac{\omega_{sp}}{2}\chi_2(\omega_{sp})\vec{E}_\omega^2(\vec{r}), \tag{12}$$

The total power absorbed will be the integral of (12) over the matrix volume

$$\int\langle\vec{E}(\vec{r})\cdot\dot{\vec{P}}(\vec{r})\rangle d^3\vec{r} = -\frac{\omega_{sp}}{2}\chi_2(\omega_{sp})\int\vec{E}_\omega^2(\vec{r})dV = -\omega_{sp}\chi_2(\omega_{sp})\frac{4\pi}{3}\frac{\vec{D}_0^2}{\varepsilon_1^2(\omega_{sp})R_0^3}, \tag{13}$$

where $R_0$ is the radius of the MNP. As the energy absorbed by the matrix is equal to the energy loss due to the damping of oscillations, i.e. $-\frac{dW}{dt}$, it follows from (1) and (13) that

$$\dot{W} = -\frac{2\chi_2(\omega_{sp})}{3\varepsilon_1^2(\omega_{sp})R_0^3\omega_{sp}}\cdot\frac{4\pi N_e e^2}{m}W. \tag{14}$$

Finally, the damping rate due to leakage of oscillations energy to the matrix is

$$\frac{1}{\tau_{res}} = \frac{2}{9}\frac{\varepsilon_2(\omega_{sp})\sqrt{1 + 2\varepsilon_1(\omega_{sp})}}{\varepsilon_1^2(\omega_{sp})}\omega_p. \tag{15}$$

As the plasma frequency of noble metals is approximately the same ($10^{16}$ rad/s) [24], the damping rate is determined mainly by the imaginary part of dielectric constant of matrix.

In other words, the stronger the absorption at SP frequency the faster the damping of oscillations. We can express $\tau_{res}$ in terms of the absorption coefficient $\alpha(\omega_{sp})$ of matrix at SP frequency

$$\frac{1}{\tau_{res}} = \frac{2}{9} c \alpha(\omega_{sp}) \frac{1 + 2\varepsilon_1(\omega_{sp})}{\varepsilon_1^{3/2}(\omega_{sp})}, \qquad (16)$$

where c is the speed of light in vacuum. For example, if $\varepsilon_1 \approx 2$ we have from (16) $\frac{1}{\tau_{res}} \approx 10^{10} \alpha(\omega_{sp}) s^{-1}$, and for $\alpha(\omega_{sp}) \sim 10^3 \, cm^{-1}$ and more this damping mechanism becomes important.

Thus, in case of resonant coupling the optical characteristics of the matrix fully determine the SP oscillations energy leakage rate from MNP.

### 3. Conclusion

Thus we have considered two possible mechanisms of the SP oscillations damping in MNP, which are not connected with electron-phonon interaction and can play main role under the conditions of strong-size quantization. The important difference between them is that the radiation damping rate manifests volume dependence, whereas the damping rate of the resonant coupling mechanism depends only on the optical characteristics of matrix. This difference will allow to distinguish experimentally the contribution of each of them into the net damping rate. While for very small MNPs with size 3-5 nm and less the radiation damping mechanism becomes ineffective (see Eq. (4)), instead there are cases when the resonance coupling provides effective energy outflow.